\RequirePackage{fix-cm}
\documentclass[smallextended]{svjour3}       
\smartqed  
\usepackage{graphicx}
\usepackage[Symbol]{upgreek}
\usepackage{amsmath}
\usepackage{hyperref}
\usepackage{dcolumn}
\usepackage{bm}
\usepackage{mathtools}
\usepackage{textcomp}
\usepackage{braket}
\usepackage{empheq}
\usepackage[Symbol]{upgreek}
\usepackage{amssymb}
\usepackage{mathrsfs}
\usepackage{dcolumn}
\usepackage{bm}
\usepackage[english]{babel}
\usepackage{hyperref}
\usepackage{times}
\usepackage{amsfonts}
\usepackage{epsfig}
\usepackage{mathrsfs}
\usepackage{color,soul}
\usepackage{bbold}
\usepackage{subfig}
\usepackage{cite}
\usepackage{pstricks}

\newcommand{\IN}[1]{\textcolor{black}{#1}}

\newcommand{\YS}[1]{\textcolor{black}{#1}}
\begin{document}

\title{Self-healing of structured light: a review}


\author{Yijie Shen, Shankar Pidishety, Isaac Nape, and Angela Dudley}


\institute{Yijie Shen and Shankar Pidishety: \at
              Optoelectronics Research Centre, University of Southampton, Southampton SO17 1BJ, United Kingdom\\
              Isaac Nape and Angela Dudley: \at
               School of Physics, University of the Witwatersrand, Private Bag 3 Johannesburg Wits 2050, South Africa\\
               Correspondence to Y. Shen:\\
\email{y.shen@soton.ac.uk}           
}
\date{Received: date / Accepted: date}

\maketitle
\begin{abstract}
\YS{Self-healing of light refers to the ability of a light field to recover its structure after being damaged by a partial obstruction placed in its propagation path.} Here, we will give a comprehensive review of the history and development of self-healing effects, especially highlighting its importance in vector vortex beams carrying spin and orbital angular momenta. Moreover, an unified zoology of self-healing, structured light is proposed to unveil a deeper understanding of its physical mechanism and provide a bird's eye view on diverse forms of self-healing effects of different kinds of complex structured light. Finally, we outline the open challenges we are facing, potential opportunities and future trends for both fundamental physics and applications.

\end{abstract}

\newpage
\tableofcontents

\newpage
\section{Introduction}
As its terminology implies, self-healing, also called self-reconstruction or self-repairing, refers to the process of recovery to the original state after disturbances, which is an effect studied as a hot topic in widespread disciplines, such as physics~\cite{macdonald1996interboard,bouchal1998self}, materials~\cite{hager2010self,wang2020self}, geophysics~\cite{lambert2021propagation}, chemistry~\cite{costentin2017self}, biology~\cite{lee2018current} and psychology~\cite{mayer2009energy}. In optics or photonics, the self-healing of light means the light field has the ability to reconstruct itself after a partial obstruction placed in its propagation path~\cite{macdonald1996interboard,bouchal1998self}. Although it seems counterintuitive, such an effect has been experimentally verified {close to 30 years ago}~\cite{macdonald1996interboard,bouchal1998self}. With the recent development of structured light~\cite{forbes2021structured,lsa}, more and more complex forms of optical self-healing effects have emerged. Whereas, the theoretical explanations to the self-healing effect airs diverse views and to date, is still in heavy debate, without a systematic summary yet. {Regardless of this and the fragmented existence of theoretical publications, optical self-healing has great promise} in applications such as optical communication, encryption, and quantum informatics. 

The development of optical self-healing largely depends on the booming {field} of structured light and the modern progress of structured light is largely based on the emergence of orbital angular momentum (OAM). In 1992, L. Allen et al. proposed that a light beam with a helical wave-front can carry OAM~\cite{allen1992orbital}, which promoted the thirty-year development of structured light to tailor patterns of amplitude, phase, and polarization of light to unveil deeper physical effects~\cite{forbes2021structured,shen2019optical}. In these thirty years, researchers have presented a large number of light beams that can carry OAM, with the most typical being the Laguerre-Gaussian (LG) beams and Bessel beams~\cite{shen2019optical}. LG beams are a form of paraxially focused eigenmodes of light beams, while the Bessel beams are nondiffracting beams{~\cite{indebetouw1989nondiffracting, mazilu2010light, vicente2021bessel}}, both of which possess an azimuthally dependent phase term of $\exp{(i\ell\phi)}$ carrying an OAM equivalent to $\ell\hbar$ per photon. Emergence of OAM light enabled the exploration of nontrivial physical effects of light, for example, the self-healing effect of light. Since the first discovery of the optical self-healing effect in Bessel beams~\cite{macdonald1996interboard}, the counterintuitive effect of light gradually became a hot topic with increasing attention. Many exotic kinds of self-healing effects were reported in general kinds of structured light in succession, especially in the OAM structured light. \YS{For instance, the self-healing was originally studied in Bessel beams with a circular symmetrical obstruction, while, recent experiments revealed that the self-healing can be observed in other mode familities too. } In addition to the experimental reports, the theoretical explanations to the striking effects, with deeper physical mechanism, were still in heated debate to date. 

In this Review, we present a comprehensive review of the history of the optical self-healing effect, from the original discovery in Bessel beams to the development towards complex structured light, including OAM beams, complex vector vortex beams, space-time structured pulses, as well as its application in both quantum and classical levels. We also review the different kinds of theoretical models for interpreting the physical origin of self-healing. Moreover, an unified zoology is proposed to provide a bird’s eye view on diverse forms of self-healing, structured light, unveiling a deeper understanding of the physical mechanism. Finally, we outline the open challenges we are still facing, potential applications, opportunities and future trends.

\begin{figure}
\centering
\includegraphics[width=\linewidth]{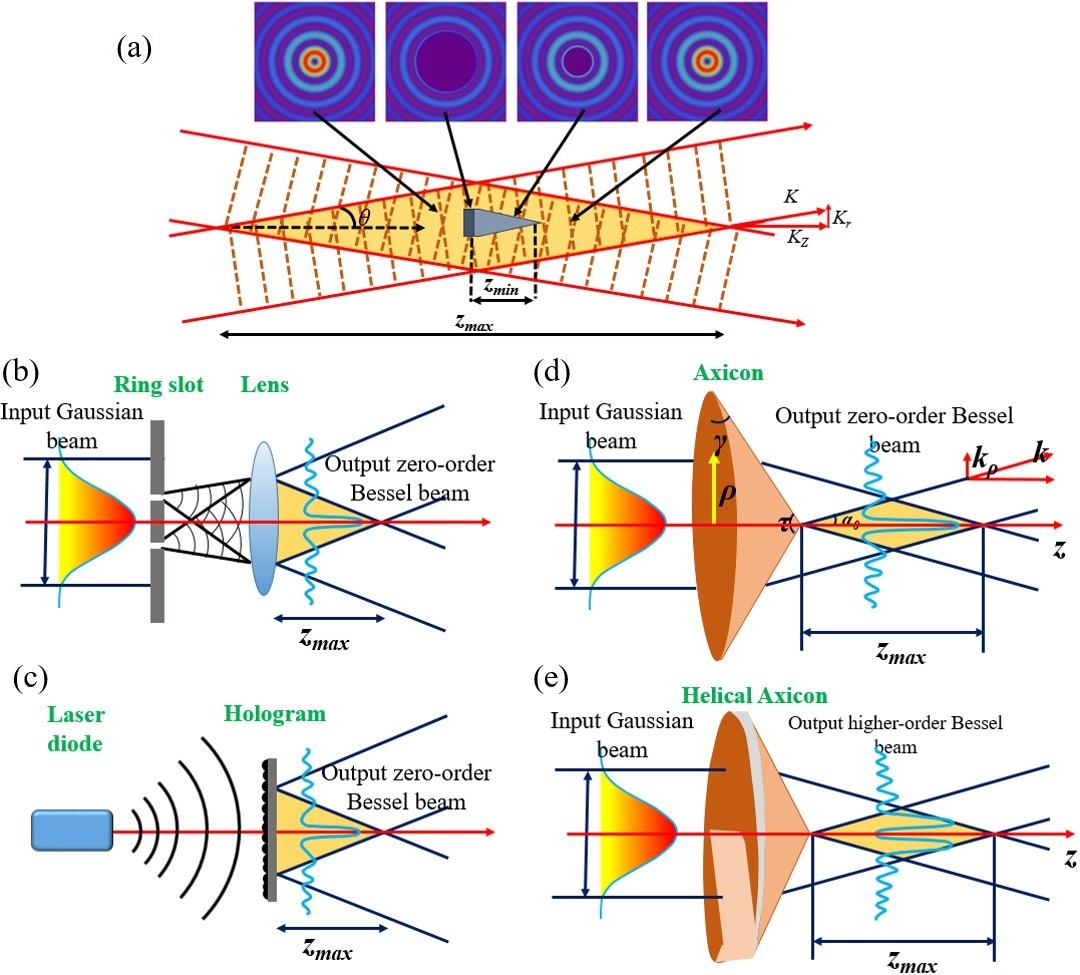}
\caption{\label{f1}{(a) A fundamental schematic of the self-healing effect of the Bessel mode. The Bessel mode exists in a finite region as the superposition of a set of planar waves propagating with the same radial momentum $k_r$, $z_{max}$ (yellow diamond). An obstacle placed in the centre of this region (black rectangle) obstructs the beam for a minimum distance, $z_{min}$ (grey triangle), after which the Bessel light field reforms. The insets display the expected image of the beam at four different planes. (b-e) The schematics of four different generation methods of Bessel beams with self-healing properties using a ring-slit and lens, laser diode and hologram, axicon, and helical axicon, respectively.}}
\end{figure}

\section{Discovery of optical self-healing}
In 1987, Durnin proposed the exact solutions for nondiffracting light beams, i.e. the Bessel beams, together with the experimental investigation, whereby the transverse pattern does not diverge upon propagation~\cite{Durnin1987,durnin1987b}. Soon after, the generation and control of such nontrivial beams became hot topics. In 1996, MacDonald et al. found an intriguing effect when executing the Bessel beam experiment, the central spot of a Bessel beam can reconstruct itself after being blocked by an obstruction, and they elaborately applied this effect into interboard optical data distribution~\cite{macdonald1996interboard}. Then, in 1998, Bouchal et al. performed deeper theoretical and experimental studies of the distorted nondiffracting beams, demonstrating that a Bessel beam, disturbed by an obstacle, is able to reconstruct its initial amplitude profile under free-space propagation, and named such effect as self-reconstruction~\cite{bouchal1998self}. Soon after, similar effects were observed and studied in more kinds of structured light, also named as self-repairing, self-recovery, and self-healing.

Here we demonstrate the basic principal of the optical self-healing effect in Bessel beams, which is the kind of structured light where such effect was first observed. Bessel beams are a class of propagation-invariant solutions to the Helmholtz equation~\cite{durnin1987b}, which can be decomposed into plane waves with wave vectors that lie on a cone~\cite{mcgloin2005bessel,vicente2021bessel}, as shown in Fig.~\ref{f1}(a) for a zero-order Bessel beam.  \IN{ A typical example, that physically resembles such a beam is the Bessel-Gauss (BG) beam \cite{gori1987bessel}. In polar coordinates $(r,\phi, z)$, BG beams can be expressed as}
\begin{align}
U_{\ell}(r,\phi, z) = &\sqrt{\frac{2}{\pi}}J_{\ell}\left(\frac{z_R k_r r}{z_R-\text{i}z}\right) \exp\left(\text{i}\ell\phi\right) \exp\left(-\text{i}k_z z\right)
\exp\left(\frac{\text{i}k_r^2z w_0-2kr^2}{4(z_R-\text{i}z)}\right),
\label{eq:BGmodes}
\end{align}
\IN{where $J_{\ell}(\cdot)$ corresponds to the Bessel function that controls the radial profile characterised by concentric rings that are determined by the transverse wavenumber, $k_r$, where $k_z$ is the longitudinal wavenumber. Furthermore, the OAM content is determined by the topological charge $\ell$, while $w_0$ corresponds to the waist size of the Gaussian envelope with a corresponding Rayleigh range $z_R = \pi w_0^2/\lambda$ for a given wavelegnth $\lambda$. The intensity profile can be seen in Fig. \ref{f1} (a)}.

Based on the conical wave picture, an opaque obstruction in the light field, \IN{ e.g., in the path of a BG beam,} will induce a conical shadow for a limited distance \IN{($z_{min}$)}, as shown in Fig.~\ref{f1}(a), and the light fields propagating before and after the shadowing region would have the same distribution. That is the most classic explanation of self-healing. Although a theoretical Bessel beam is theoretically nondiffracting in an infinite propagation distance~\cite{Durnin1987,durnin1987b}, a physical beam generated from an actual experiment only exists for a limited interference distance~\cite{lapointe1992review}. This effective region is also called Bessel beam shadowing. 

{Diverse generation schemes were proposed to experimentally generate Bessel beams. These consist of refractive optical elements, for example axicons~\cite{scott1992efficient, herman1991production, mcleod1954axicon }, and diffractive optical elements in the form of slit diffraction~\cite{Durnin1987} and computer generated holograms~\cite{turunen1988holographic, vasara1989realization, davis1993diffraction, davis1996nondiffracting, davis1996intensity, paterson1996higher}, refer to Figs.~\ref{f1}(b-d). All of the aforementioned techniques can be used or adapted to generate both zero and higher-order Bessel beams. Zero-order Bessel beams possess a bright central maximum, while higher-order Bessel beams contain a central null (or vortex), that propagates in a nondiffracting manner. These higher-order Bessel beams carry OAM owing to their azimuthal phase variation~\cite{chavez1996nondiffracting}. With the advancement in 3D printing and manufacturing, higher-order Bessel beams can be generated by customized helical axicons~\cite{wei2015generation, topuzoski2009conversion}, see Fig.~\ref{f1}(e). Extending the aforementioned generation techniques even further, superpositions of higher-order Bessel beams can also be generated, consisting of - illuminating Durnin’s ring-slit aperture with multiple azimuthal phase components~\cite{vasilyeu2009generating} or illuminating an axicon with a superposition of LG beams~\cite{arlt2000generation, rop2012measuring}.
While, there are many ways to generate Bessel beams, it is worth noting that some recently advanced methods, such as digital holograms~\cite{dudley2013unraveling} and ring-shaped (annular) lenses~\cite{vetter2019realization}, have already made it possible to create long-distance self-healing Bessel beams~\cite{ belyi2010bessel, litvin2015self}.}


With the development of manipulating structured light, more and more types of complex structured modes can be precisely controlled~\cite{forbes2021structured,forbes2020structured}. This has subsequently led to the self-reconstruction effect being demonstrated with other forms of structured light, namely, higher-order Bessel vortex, Airy, and caustic modes, to name but a few. 

\begin{figure}
\centering
\includegraphics[width=1.4\linewidth]{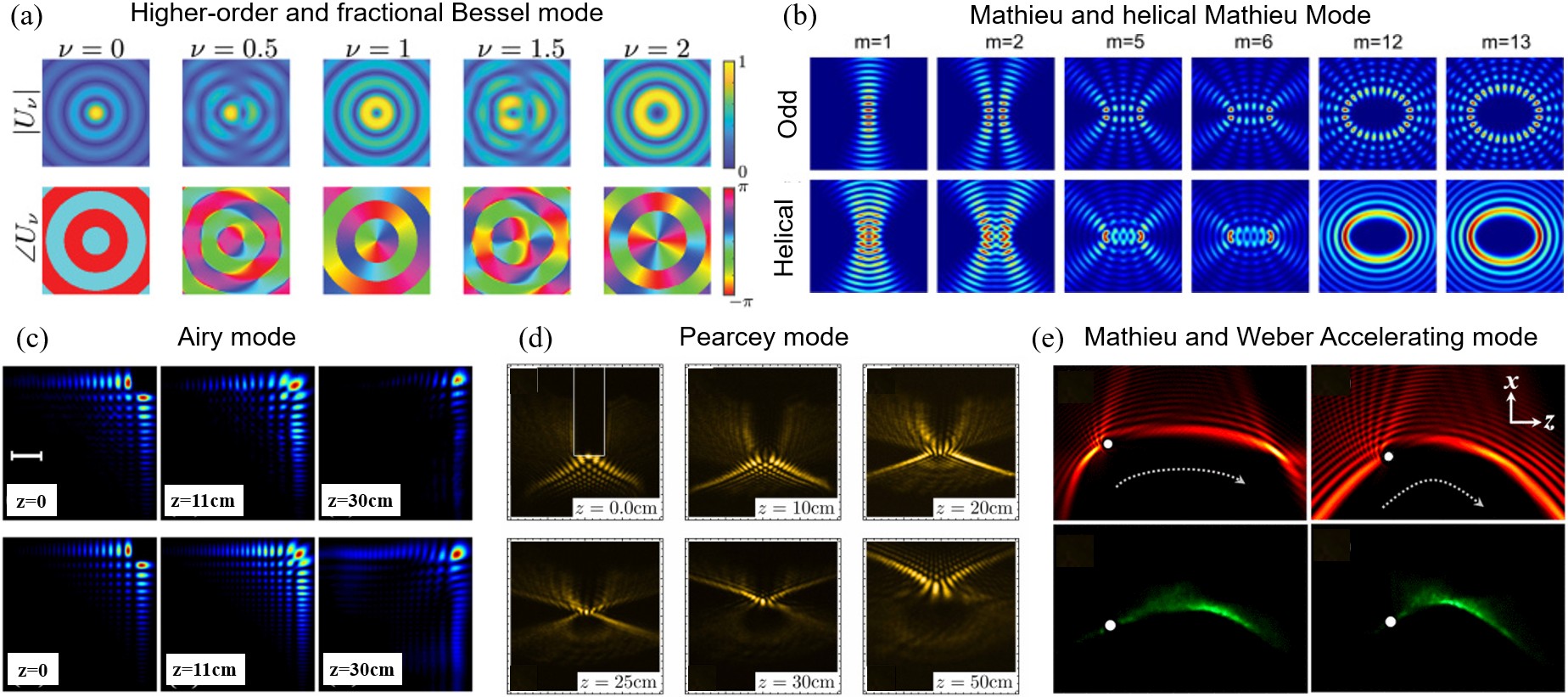}
\caption{\label{f2}{Various kinds of self-healing modes: (a) amplitude and phase distributions from fundamental-order to higher-order and fractional Bessel modes carrying OAM~\cite{vicente2021bessel}, \YS{Copyright by Optica Publishing Group.}; (b) Intensity distributions of a selection of odd and helical Mathieu beams~\cite{alpmann2010mathieu}, \YS{Copyright by Optica Publishing Group.}; (c) Theoretical and experimental results of self-healing Airy beams upon propagation~\cite{broky2008self}, \YS{Copyright by Optica Publishing Group.}; (d) \YS{Experimental} results of self-healing Pearcey beams upon propagation~\cite{ring2012auto}, \YS{Copyright by Optica Publishing Group.}; (e) Theoretical and experimental results of self-healing Mathieu and Weber accelerating modes upon propagation~\cite{zhang2012nonparaxial}, \YS{Copyright by American Physical Society.}}}
\end{figure}

\section{Basic models of optical self-healing}
Since optical self-healing was discovered in Bessel beams, an open question has arisen, namely: what is the basic mechanism for self-healing? Is the self-healing a unique phenomenon in diffraction-free optical fields? In this chapter, we review diverse kinds of explanations to the self-healing effect together with corresponding experimental verification in a historical order, which renews our understanding for self-healing.

\subsection{Nondiffraction explanation}
For a long time, people believed the nondiffractive nature is the main mechanism to induce self-healing, in other words, self-healing was considered as a distinctive feature of nondiffraction. Theoretically, Bessel beams are exact solutions to the expressed Helmholtz equation in longitudinally-independent cylindrical polar coordinates~\cite{durnin1987b}. While, longitudinally-independent spatial coordinates can be selected as other forms of transverse coordinates, such as parabolic and elliptical coordinates to solve the nondiffracting structured beams. Thus, the the family of nondiffracting beams includes many other general modes in addition to the standard Bessel-type, such as fractional Bessel modes~\cite{vicente2021bessel} (Fig.~\ref{f2}(a)), Mathieu modes~\cite{chavez2001elliptic} (Fig.~\ref{f2}(b)), parabolic modes~\cite{bandres2004parabolic}, and other general nondiffracting structured modes~\cite{lopez2010shaped,thomson2008holographic,ren2021non}.

After the verification of self-healing in zero-order Bessel beams, this too was noted in many other kinds of nondiffracting modes, such as higher-order Bessel beams carrying OAM~\cite{bouchal2002resistance,hu2020long,yang2018anomalous}, fractional Bessel beams~\cite{tao2004self,ehsan2020space,vicente2021bessel}, asymmetric Bessel beams~\cite{anguiano2018self}, parabolic modes~\cite{lopez2005observation,ruelas2013engineering}, and Mathieu beams~\cite{alpmann2010mathieu}. In these experimental explorations, various kinds of nondiffracting modes were generated and self-healing properties were verified. However, an ideal nondiffracting beam over infinite distance does not physically exist. One can only study the properties in an effectively confined region (as shown in Fig.~\ref{f1}(a)). Alternatively, one can use a pseudo-nondiffracting beam to study self-healing, which is obtained by modulating a nondiffracting beam with a Gaussian envelope, also termed elegant Gaussian modes, {and in the case of the Bessel function, as BG beams ~\cite{gori1987bessel}}. In contrast to standard nondiffracting beams, the pseudo-nondiffracting beams can be experimentally generated with high fidelity within a full propagation domain. The self-healing properties of these kinds of pseudo-nondiffracting beams were also fully explored~\cite{chabou2020elegant,bencheikh2020cosine}, and can be explained by their nondiffracting nature, while the Gaussian modulation has no impact.  

\begin{figure}
\centering
\includegraphics[width=1.4\linewidth]{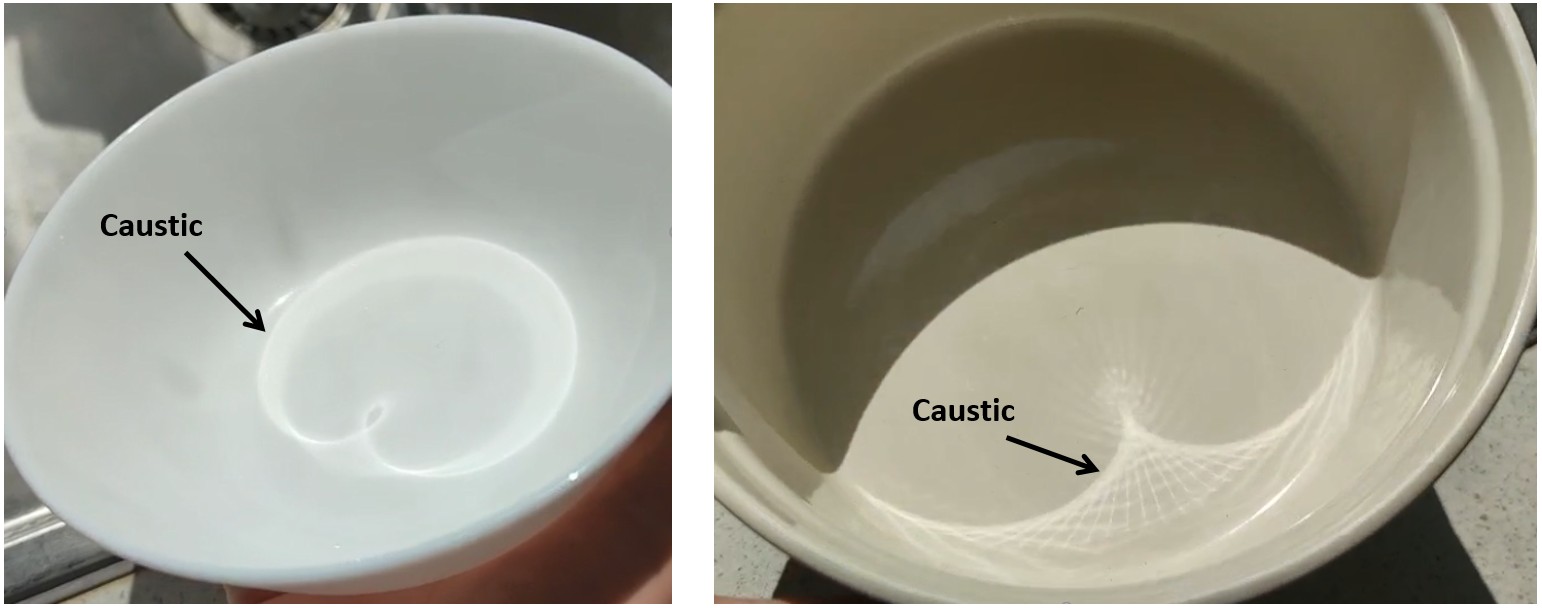}
\caption{\label{f3}{Caustics in daily life: diversified caustics can be observed in the bottom of bowls under natural light illumination. Photoed by Yijie Shen.}}
\end{figure}

\begin{figure}
\centering
\includegraphics[width=\linewidth]{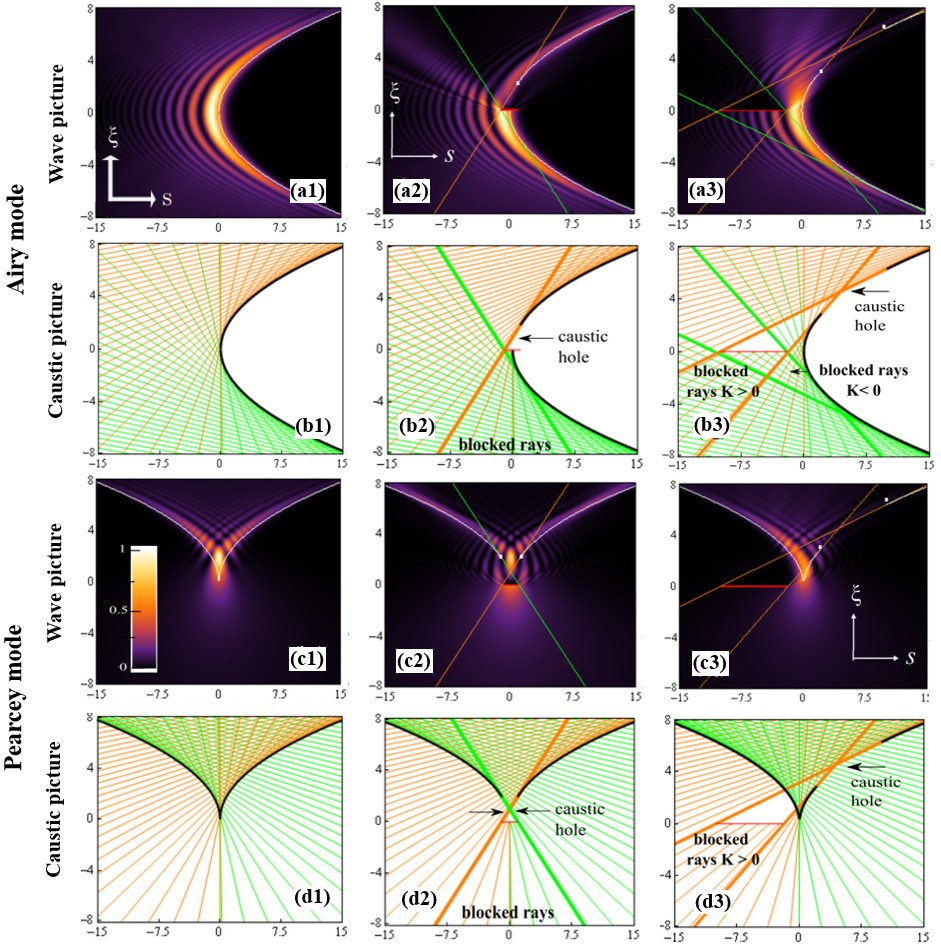}
\caption{\label{f4}{Caustics in explaining the self-healing effects and characterizing the strength for an Airy mode and a Pearcey mode~\cite{vaveliuk2017dual}: (a1) The intensity distribution of a propagating Airy mode, (a2) the Airy mode obstructed by a small obstacle, and (a3) the Airy mode obstructed by a larger obstacle. (b1-b3) The caustic interpretations corresponding to the self-healing modes (a1-a3). (c1) The intensity distribution of a propagating Pearcey mode, (c2) the Pearcey mode obstructed by a small obstacle, and (c3) the Pearcey mode obstructed by a larger obstacle. (d1-d3) The caustic interpretations corresponding to the self-healing modes (c1-c3). \YS{Copyright by American Physical Society}}}
\end{figure}

\subsection{Ray/Caustic explanation}
After the emergence of many studies pertaining to self-healing in nondiffracting beams, researchers realized that nondiffraction is not a sufficient condition for self-healing, because many self-healing effects were demonstrated in other types of structured light. {Instead, the self-healing could be simply modeled via geometric optics~\cite{litvin2009conical}.} {Often, self-healing modes encompass self-accelerating modes}, such as Airy~\cite{chu2012analytical,anaya2021airy,broky2008self} (Fig.~\ref{f2}(c)), Pearcey~\cite{ring2012auto} (Fig.~\ref{f2}(d)), accelerating parabolic~\cite{bandres2008accelerating}, accelerating Weber and accelerating Mathieu beams~\cite{zhang2012nonparaxial,bandres2013nondiffracting} (Fig.~\ref{f2}(e)). 
The intensity pattern of such a mode does not evolve perfectly unchanged upon propagation, as in the case of a nondiffracting beam, but instead evolves {unchanged} along an accelerating trajectory. Therefore, the standard nondiffracting modes can be seen as an extreme case of the accelerating modes~\cite{bandres2009accelerating,bandres2013accelerating,yan2015accelerating,julian2018wavefronts}. Since the accelerating modes have salient trajectory characteristics, a ray-coupled theory, i.e. caustic theory, was recently developed to model nondiffracting structured modes with unified representation of general geometric patterns~\cite{zannotti2020shaping}. Since the theory of caustic beams is established, innumerable links between caustics and self-healing properties were found~\cite{anguiano2007self}. 

What are caustics? A caustic is the envelope of light rays reflected or refracted by a curved surface or object, which can form ruled ray-coupled geometric patterns. Actually, caustics can be seen in our daily light, e.g. the Nephroid caustic which appears at the bottom of a bowl after natural light illumination, see Fig.~\ref{f3}. Caustics are also used as effective tools to tailor structured light. The caustic structured light, including nondiffracting and self-accelerating beams, have an interesting property in that the wave-packets are always coupled with caustics~\cite{forbes2001using}. Subsequently, the caustic has become an endorsed tool to explain the basic mechanism of self-healing~\cite{vaveliuk2017dual}, and the previous nondiffraction and self-accelerating beams are merey special forms of caustic beams. Caustics can effectively predict the self-healing region in corresponding structured beams with a given obstacle. For instance, the caustic representations of an Airy beam and a Pearcey beam are shown in Figs.~\ref{f4}(a,b) and \ref{f4}(c,d), respectively. With an obstacle set in a caustic beam, the blocked rays can be identified and the region with absence of interlaced rays, termed the caustic hole, can be easily found (Fig.~\ref{f4}). Such caustic holes can precisely characterize the range and strength of self-healing.

\begin{figure}
\centering
\includegraphics[width=1.2\linewidth]{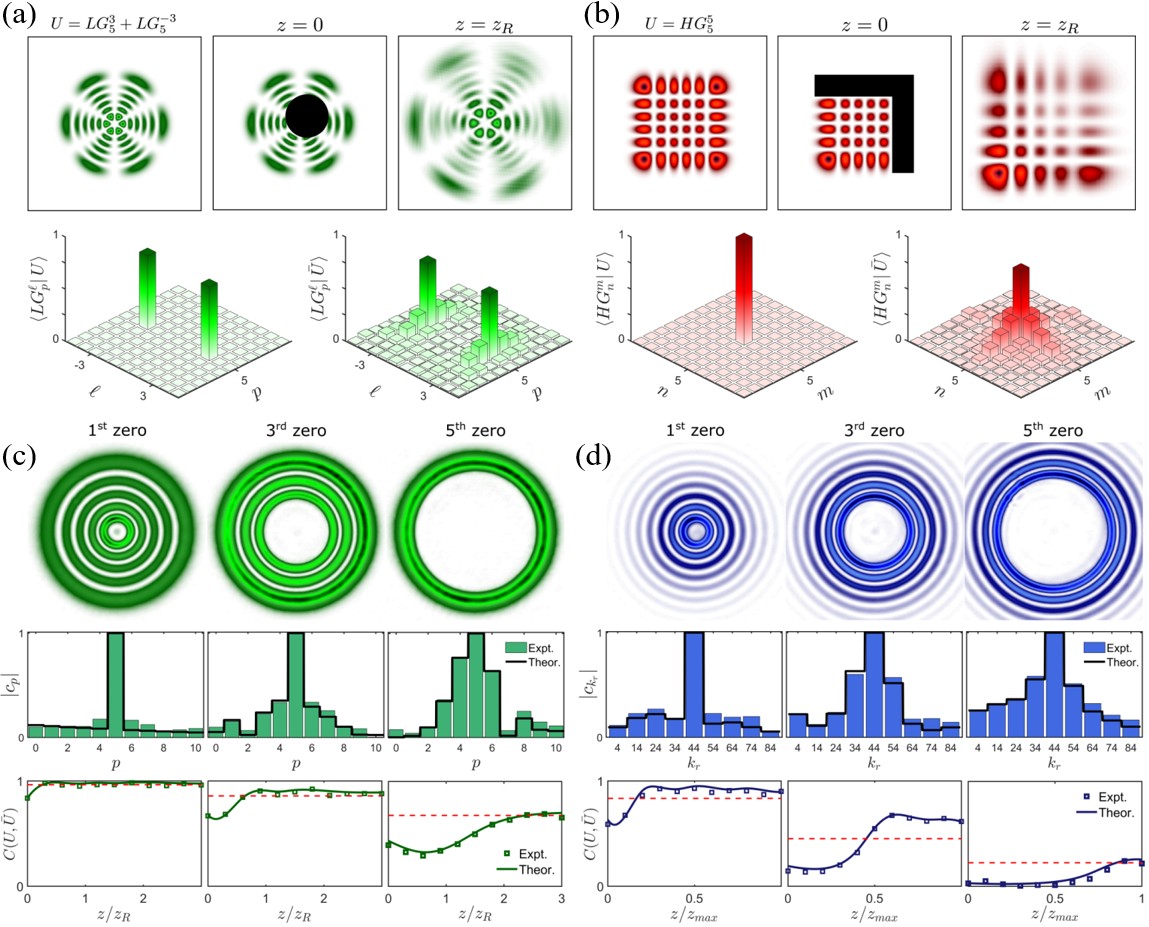}
\caption{\label{f5}{(a,b) Self-healing properties of a conjugate-symmetric LG mode and a HG mode, together with the analysis of modal decomposition. Whereby, the obstructed mode after propagation can have a measured modal decomposition matrix very close to the original (second row). For the LG (HG) beam, the field amplitude is maximally self-reconstructed by 94.8\% (64.6 \%) at the plane $z=0.6z_R$ ($0.8z_R$), with a mode spectrum fidelity of 0.81 (0.50). For this particular example, there is a clear link between the mode spectrum fidelity and the degree of self-healing. (c,d) Modal analysis of the self-healing processes of a LG mode and a Bessel mode, versus different obstacle sizes and propagation distances. Shown are the experimental beam images for different sized obstructions (first row), the corresponding rescaled radial mode spectra (second row), and the field amplitude correlation as a function of propagation distance (last row). Dashed lines in the last row correspond to the modal spectrum fidelities with the unobstructed beam. Experiment (Expt.) and theory (Theor.) are in excellent agreement~\cite{pinnell2020modal}. \YS{Copyright by American Physical Society}}}
\end{figure}

\subsection{Wave explanation}
Another direct interpretation of self-healing of various optical modes is using the classic diffracting wave-optics simulation, for a typical example, the self-healing of optical vortex beams.
An optical vortex refers to a beam with helical phase and a phase singularity with a certain topological charge, which reveals the amount of OAM the beam carries~\cite{shen2019optical}. It was experimentally demonstrated that an optical vortex can possess the self-healing property. If one uses an obstacle to block the phase singularity of a vortex beam, after propagation of a certain distance, the phase singularity with the same topological charge can be revived in the beam~\cite{vasnetsov2000self}. Many theoretical simulations based on diffractive optics were made to describe the self-healing process with details ensuing their experimental discovery~\cite{bekshaev2014transverse,bekshaev2017singular}. In addition, the self-healing properties of optical vortices were extended to higher-order complex vortex modes with large topological charges~\cite{bekshaev2015spatial}. 

\begin{figure}
\centering
\includegraphics[width=1.3\linewidth]{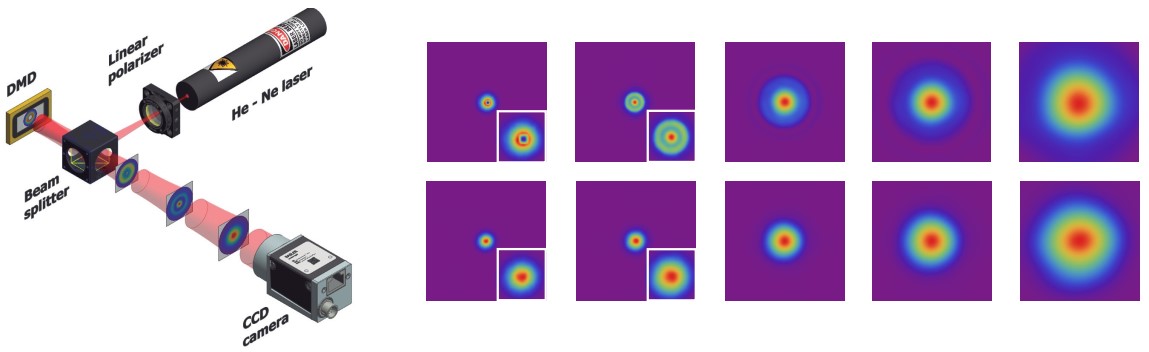}
\caption{\label{f6}{Demonstration of the self-healing effect of a fundamental Gaussian beam: setup (left) and results of obstructed and original beams (right upper and lower)~\cite{aiello2017unraveling}. \YS{Copyright by Optica Publishing Group}}}
\end{figure}

Inspired by the study of the self-healing of optical vortices, similar studies for other general kinds of optical wave modes based on diffractive modeling were also reported, for instance, the self-healing of LG, Hermite-Gauss (HG) and Ince-Gauss (IG) beams~\cite{aguirre2015self, ugalde2021traveling, xu2020self, zhao2021self,shen2018beam,pinnell2020revealing}. The wave-optics description of the self-healing mechanism is also suitable in explaining a Bessel beam's self-healing in a quantitative manner~\cite{aiello2014wave,chu2014quantitative}. Interestingly, a recent study has demonstrated that some common eigenmodes, such as HG or LG modes, can possess an improved self-healing effect than that possessed by Bessel modes~\cite{pinnell2020revealing}. Exploiting the analytical method of modal decomposition, the construction and quality of arbitrary structured modes can be quantitatively measured~\cite{pinnell2020modal}, which provides an effective tool to compare the strength of self-healing effects of diverse kinds of structured light. For instance, by comparing the modal decomposition matrices of the original mode and the obstructed mode by determining their 'closeness' (or mode spectrum fidelity), see Fig.~\ref{f5}(a,b), where the resulting fidelity can evaluate the self-healing strength, it can be shown that (in some cases) the LG self-healing is better than the Bessel self-healing (Fig.~\ref{f5}(c,d)). \YS{In addition to the self-healing effects in HG and LG modes, more interestingly, similar phenomenon was also observed even in a fundamental Gaussian beam,} where it was surprisingly found that even a very common Gaussian beam can possess self-healing effect~\cite{aiello2017unraveling}, see Fig.~\ref{f6}. 

\IN{In demonstrating this the authors model the situation as an eigenvalue problem of the field function, i.e,}
\begin{equation}
    \Psi_O(r, \phi, z) \approx  \lambda_O \Psi(r, \phi, z=0),
\end{equation}
\IN{where $\Psi_O(r, \phi, z)$ is the field,  $\Psi(r, \phi, z=0)$, after it interacts with an opaque obstruction and is propagated to the position $z$ in the longitudinal direction and $\lambda_O$ is the corresponding eigenvalue. This equations holds for $z \geq z_{min}$ where $z_{min}$ and the distance after which the field should be reconstructed. Remarkably, in the fourier domain a similar eigenvalue equation can be derived, where the $z$ dependence disappears, showing that fields that are most likely to self heal are those that have angular spectra that remain unchanged by the pertubation. The authors also introduce a new distance measure that instead quantifies the degree of self healing arcording to a metric that measures the closeness between the intial and perturbed beam. The method was tested experimentally on Gaussian beams and produced results that are in agreement with the theory.}

Therefore, contrary to popular belief, the self-healing effect is not specific for nondiffracting and caustic beams, and instead can exist for very general free-space eigenmodes of the wave equation, including HG mode, LG mode, IG mode, and even the fundamental Gaussian mode.

\begin{figure}
\centering
\includegraphics[width=1.1\linewidth]{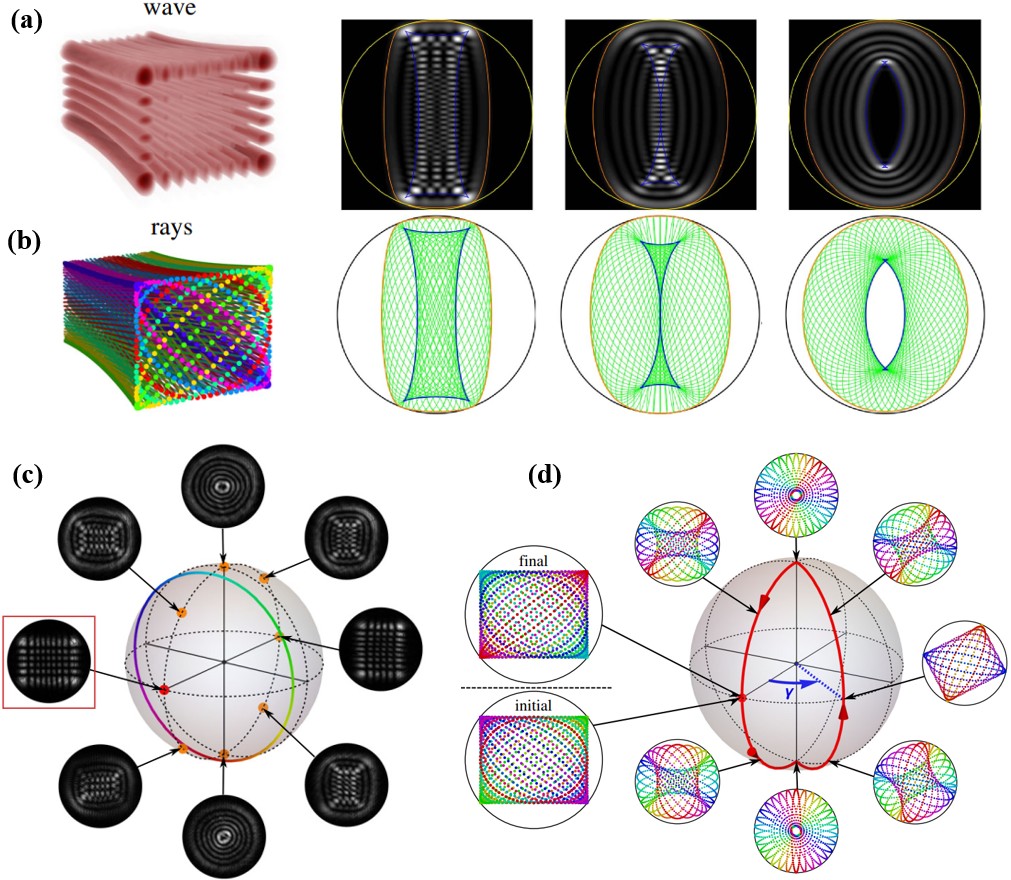}
\caption{\label{f7}{(a,b) The wave and ray treatments for the same HLG beams. (c) The modal Poincar\'e sphere representation for interpreting the evolution of a higher-order HLG mode. (d) The ray optical Poincar\'e sphere representation for the corresponding HLG mode, which also can reveal the geometric phase evolution without interferometry~\cite{malhotra2018measuring,alonso2017ray}. \YS{Copyright by American Physical Society}}}
\end{figure}

\subsection{Ray-wave explanation}
Treated by ray and wave representations respectively, the self-healing of light can be interpreted by different theoretical models. Whilst, the ray-wave duality treatment in explaining unified effects of structured light has become a hot topic in recent years~\cite{shen2021rays}. Although, the wave description predicted that any free-space eigenmode can have a self-healing effect, this seems far beyond what the ray representation can describe. However, with the development of ray-wave theory, a unified description can be established, because the free-space wave eigenmodes can also be treated as ray-like. For a typical example, the Hermite-Laguerre-Gaussian (HLG) modes were always seen as the general structured Gaussian eigenmodes under astigmatic transformation~\cite{abramochkin2004generalized,abramochkin2010general,shen2019hybrid}, see Figs.~\ref{f7}(a,b). The complete evolution of which can be mapped onto a modal Poincar\'e sphere~\cite{dennis2017swings,shen20202,gutierrez2020modal}, see Fig.~\ref{f7}(c). In parallel with the structured Gaussian modes, a ray-optical Poincar\'e sphere was proposed, which describe the transformation of a set of ray caustics~\cite{alonso2017ray}, see Fig.~\ref{f7}(d). Importantly, the caustic ray transformation is exactly coupled with the corresponding HLG wave modes~\cite{dennis2019gaussian}. Based on the ray-wave duality, it is no longer surprising that common wave eigenmodes can be self-healing, apart from the special nondiffracting and accelerating modes, as the caustic representation is still available in widely explaining the wave-optical effect. For instance, the caustic representation can provide a geometric ray interpretation of the geometric phase, which shows great agreement with the conventional wave interference interpretation~\cite{malhotra2018measuring}. On the other hand, the corresponding caustic representations for more kinds of classical wave modes were exploited~\cite{alonso2021abstract}. For instance, the set of ray caustics representing IG modes was proposed~\cite{gutierrez2021emulating}, that can be a perfect explanation as to why the IG modes can be self-healing. In summary, the ray-wave duality methodology provides a unified and deeper perspective in explaining the self-healing effect in more general types of structured light.

\subsection{\IN{Impact on photon states}}
\IN{To extend the explanation of self-healing to the single photon regime, we take a step back and remind the reader of the simple wave equation in the paraxial regime and show how self healing of photons can be understood from simple wavetheory, following \cite{sorelli2018diffraction}. This is because self-healing should hold in the quantum  regime since quantum mechanics is essentially a wave theory.}

In the cartesian coordinates, $(x,y,z)$, the  field profile, $\psi\left(x, y, z\right)$ for a monochromatic beam propagating in the longitudinal direction ($z$) satisfies the paraxial Helmholtz equation
\begin{equation}
    2ik\partial_z \psi\left(x, y, z\right) = \nabla_T^2 \psi\left(x, y, z\right),
    \label{eq:waveequation}
\end{equation}
where $\partial_z$ is the partial derivative in the longitudinal direction and $ \nabla_T^2 = (\partial^2_x, \partial^2_y)$ is the Laplacian in the transverse plane. It can be seen that the wave equation also bares resemblance to the Schrödinger equation, where the longitudinal component replaces time. It is not surprising that single photons also satisfy the same equation; an identical equation can be derived for multimodal photon states when the electric fields and field potentials in Maxwell's equation are replaced with their operator equivalent \cite{grynberg2010introduction}. Consequently, at the single photon level, the transverse amplitudes inherit similar features as mode fields that satisfy the classical wave equations, and self-healing is one of those salient properties.  We explore some nonclassical features of self-healing, e.g. entanglement self-healing  in two photon systems, and show that they can be explained by simple intermodal modal overlaps in single photons, as shown in present literature.

To begin, Eq. (\ref{eq:waveequation}) can be solved in cylindrical coordinates and admits solutions with characteristic azimuthal profiles, $\exp(i \ell \phi)$, where each photon carries discrete amounts of OAM of $\pm \ell \hbar$ for $\ell \in \mathbb{Z}$ \cite{allen1992orbital}. These photon states also form a basis for twin photons generated via spontaneous parametric down conversion \cite{mair2001entanglement} due to OAM conversation in the SPDC process.  Bessel-Gaussian (BG) modes have been shown to form a basis for SPDC states \cite{mclaren2013two}. For two photons entangled in the  OAM basis and having a self-healing BG radial profile, the state can be expressed as \cite{mclaren2012entangled}
\begin{equation}
    \ket{\Psi} = \sum_{\ell} \iint a_\ell (k_{r_1}, k_{r_2}) \ket{\ell, k_{r_1}} \ket{-\ell, k_{r_2}} dk_{r_1}, dk_{r_2}.
    \label{eq:twophotonstate}
\end{equation}
Here $|a(k_{r_1}, k_{r_2})|^2$ is the joint probability for detecting two photons in the state  $\ket{\pm \ell, k_{r_{1,2}}}$. In Ref. \cite{mclaren2012entangled} the authors showed that SPDC photons can be entangled in the BG basis and confirmed this by reconstructing the density matrix of subspaces with dimensions $d=2$ reaching up to $d=6$ dimensions with state fidelities ranging from $F_2 = 0.96 \pm 0.01$ to $F_6 = 0.79 \pm 0.01$. The next step was to show that the entanglement is robust against solid obstructions. Indeed, this was reported in \cite{mclaren2014self}.

An in-depth analysis of the self-healing property at the single photon level was reported by Sorelli et al. \cite{sorelli2018diffraction}. The authors employed a Fourier optical approach to deriving the analytical equations for the decay of the entanglement between two photons. To achieve this, they computed the scattering coefficients given a photon that encounters an obstruction. The mapping can be summarised as 
\begin{equation}
    \ket{\ell, k_{r_1}} \rightarrow \int c_{\ell}(z,k_r) \ket{\ell, k'_{r}} dk'_{r},
\end{equation}
where $c_{\ell}(z, k_r)$ are the scattering coefficients at a distance $z$ from the plane where the obstruction and incident photon field overlap. Accordingly, the scattering coefficients were computed from \cite{sorelli2018diffraction}
\begin{equation}
c_{\ell}(k_r) = \iint u^*_{BG}(x, y, z) \psi_{BG}(x, y, z) dx dy,
\end{equation}
where $u^*_{BG}(x, y, z)$ is the $z$ dependent unperturbed field profile while $\psi_{BG}(x, y, z)$ is the perturbed field profile that can be computed from the angular spectrum method. The overlap probabilities are independent of the $z$ translation, meaning that the modal scattering is only dependent on the overlap between the perturbed and unperturbed field at the initial plane. While the azimuthal component remains unaffected when the obstruction is centred with the object, the detection probability is sensitive to the relative displacement of the opaque translation in the transverse plane.



\section{Zoology of complex self-healing effects}
In the {previous} chapter, we reviewed the diverse explanations for the fundamental optical self-healing effects, which are based on classical scalar light beams. Structured light also includes more general forms, such as vector beams, spatiotemporal waves, and quantum photon states. Here, the self-healing effects are widely discussed in these complex forms of structured light.

\begin{figure}
\centering
\includegraphics[width=1.2\linewidth]{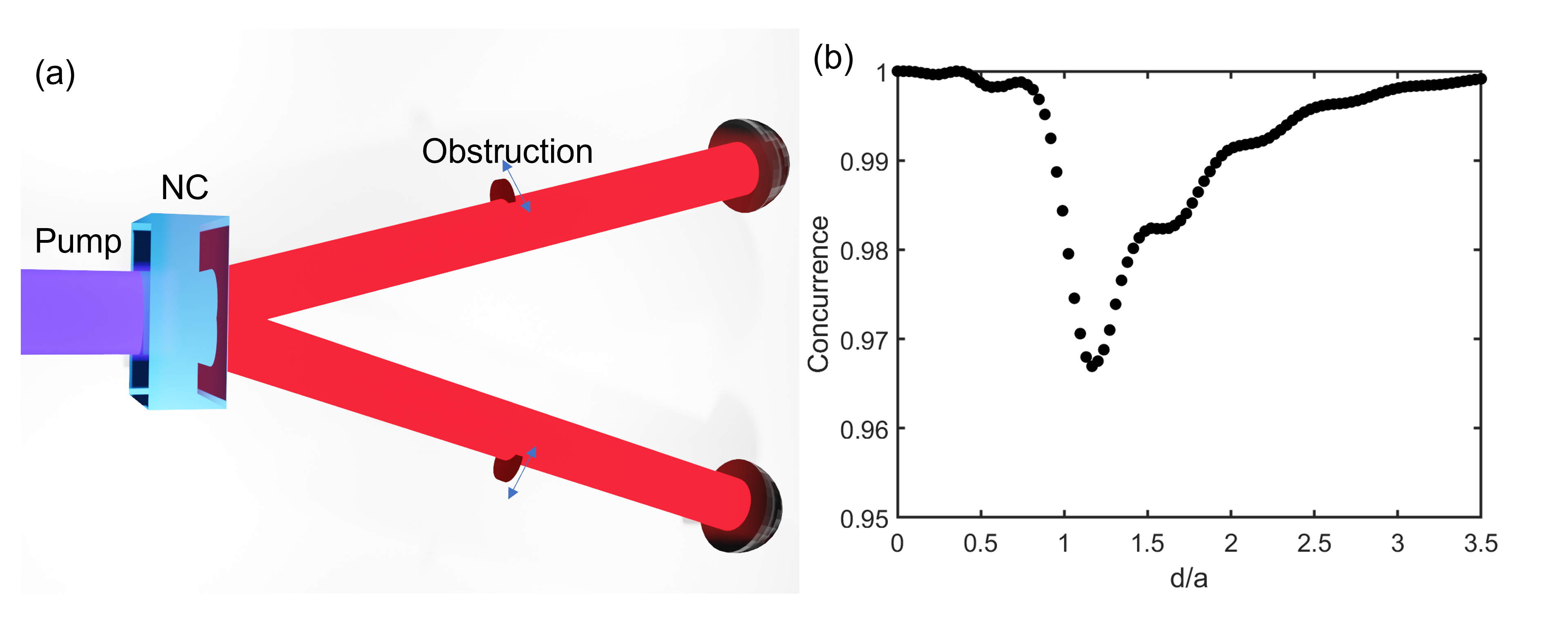}
\caption{\label{f8}{Self-healing of Bessel mode entangled photon pair: (a) \IN{a typical setup for generating self-healing photons using nonlinear crystals that are pumped with high energy photons similar to \cite{mclaren2014self}; Theoretical plot of the concurrence (degree of entanglement) vs $d/a$, where $d$ is the obstruction displacement and $a$ is the radius of the obstruction. The formulae for computing the concurrence is derived in \cite{sorelli2018diffraction}.} }}
\end{figure}

\subsection{Quantum Self-healing}
The self-healing effect has not only been studied in classical structured light, but it has also been exploited in quantum optics, i.e. by investigating the self-healing of entangled photon states~\cite{georgescu2014entanglement}. Quantum entangled states between photons can be constructed not only with the polarization state of light, but also with the spatial mode, for instance, photons can be entangled in their OAM state~\cite{mair2001entanglement,erhard2018twisted}. {Bessel modes can also be engineered to possess OAM, whereby their quantum entanglement and self-healing ability (at the single-photon level) has been verified~\cite{mclaren2012entangled}.}
Conventionally, quantum entanglement between photon pairs is fragile and can be easily destroyed by losses and noise in media. While, self-healing entangled photon pairs can overcome this drawback. When observing entangled Bessel photon pairs, the presence of a partial obstruction introduces losses that mask the correlations associated with entanglement, thus the entanglement can be revived after propagation beyond the obstruction \IN{as reported in \cite{mclaren2014self}}. \IN{Sorelli et al. \cite{sorelli2018diffraction}, showed that self healing Bessel mode entanglement is also robust against transverse translations of the obstructions, obtaining concurrences above 0,95 as shown in Fig. \ref{f8}}. \IN{Moreover, the  theoretical \cite{sorelli2018diffraction} and experimental \cite{mclaren2012entangled} evidence suggests that}  Bessel-based self-healing quantum entanglement is more robust to the losses in optical path than that in the usual LG based OAM entanglement. In addition to the self-healing of quantum entanglement, the self-healing modes of a single photon state were also studied~\cite{li2021self}. The quantum self-healing effects can overcome some limitations of quantum state distribution in the transmission path, which has been applied in the quantum protocol such as high-dimensional quantum key distribution~\cite{nape2018self}, offering advantages in more secure and robust quantum communication.

\begin{figure}
\centering
\includegraphics[width=0.7\linewidth]{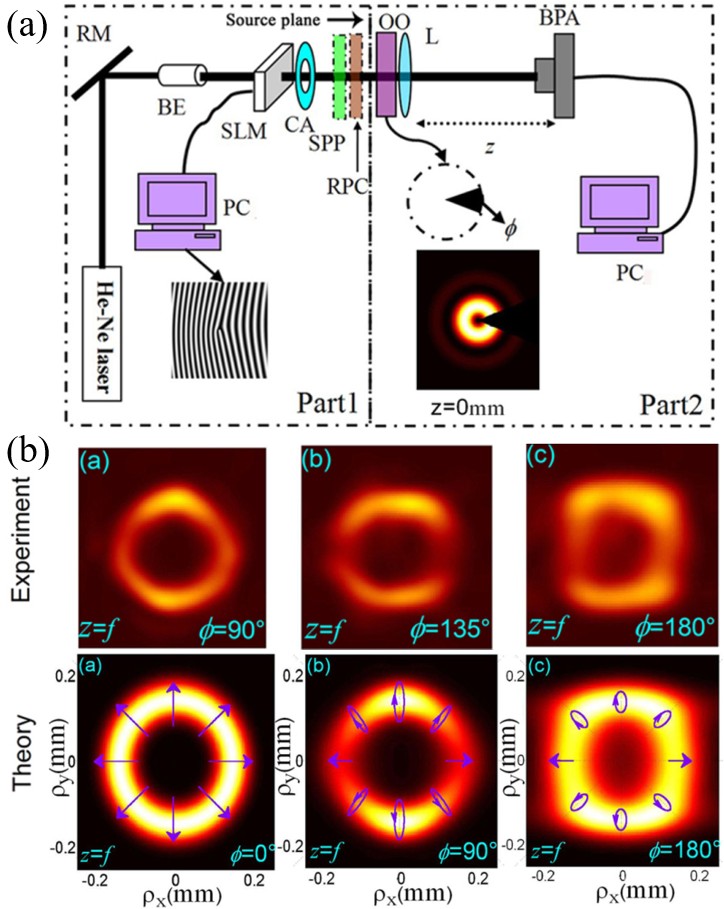}
\caption{\label{f9}{Self-healing of vector beams: (a) setup; (b) experimental results of self-healed radially polarized vector beams~\cite{wu2014generation}. \YS{Copyright by American Physical Society}}}
\end{figure}

\subsection{Self-healing of vector beams}
The self-healing effect has not only been studied with scalar fields, but also extended to vector beams. In contrast to common beams with a certain state of polarization, vector beams possess patterns of spatially variant polarization, such as radially polarized beams or azimuthally polarized beams~\cite{zhan2009cylindrical}. Such space-polarization non-separability is extremely similar to the non-separability of quantum entangled photon pairs, widely termed classical entanglement, thus many quantum analogue methods can be exploited to characterize the properties of vector beams~\cite{forbes2019classically,shen2021creation,https://doi.org/10.1002/lpor.202100533}. {Vector beams have gained interest, due to their ability to produce tight focal spots when focused by high numerical aperture objectives~\cite{brown2011unconventional}.}

\begin{figure}
\centering
\includegraphics[width=1.2\linewidth]{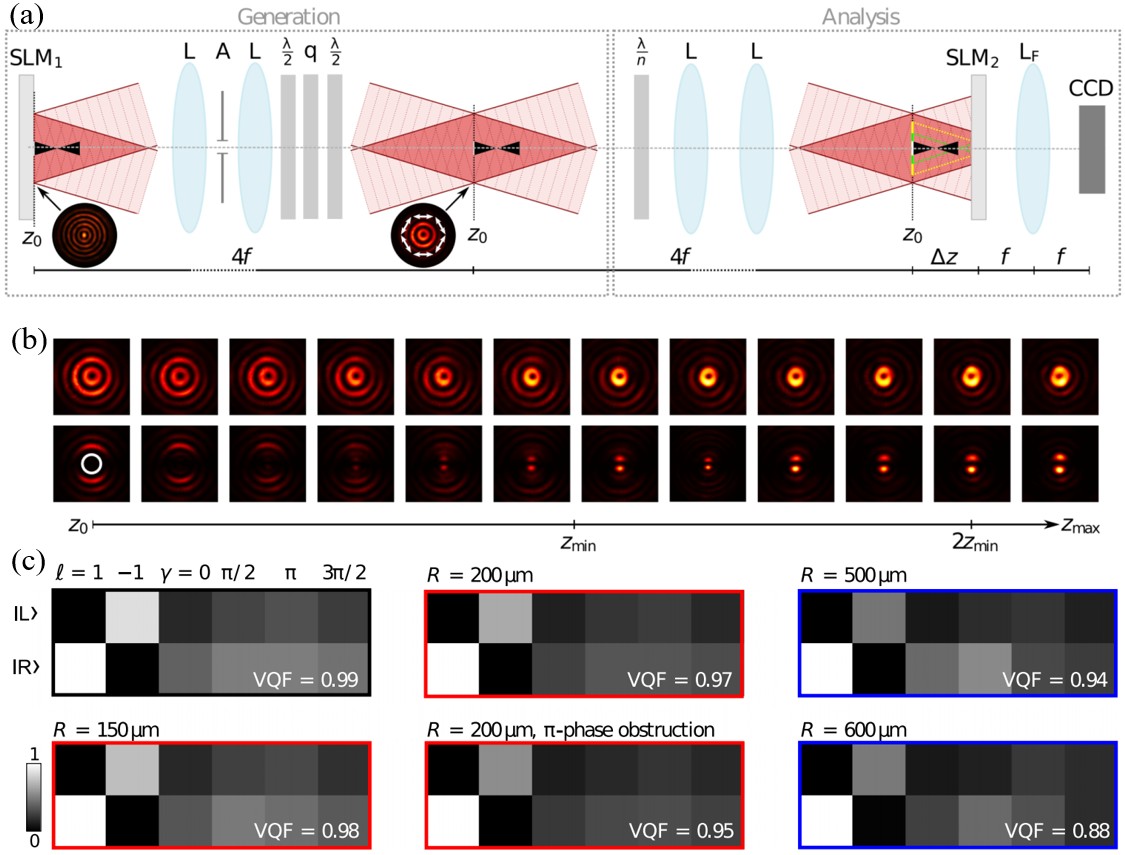}
\caption{\label{f11}{Self-healing of classical entanglement: (a) setup; (b) experimental results of self-healing vector beams; (c) experimental results of self-healing classical entanglement~\cite{PhysRevA.98.053818}. \YS{Copyright by American Physical Society}}}
\end{figure}

{Not only can Bessel beams be scalar in nature (spatially homogeneous) bu they can also have vector states of polarization (spatially inhomogeneous)~\cite{bouchal1995non,ornigotti2013radially}.} The self-healing properties in vector beams was first studied in a radially polarized BG beam, i.e. the superposed mode of two BG beams with opposite OAM and orthogonal polarizations, which were observed both in paraxial~\cite{wu2014generation,milione2015measuring} and tightly focused conditions (Fig.~\ref{f9})~\cite{vyas2011self,vyas2011diffractive}. Whereby, a radially polarized vector pattern can reconstruct itself upon propagation after a partial obstruction. The self-healing of vector beams was soon after verified for many other kinds of vector states, such as the variant states between radially and azimuthally polarized modes~\cite{li2017generation}, the higher-order-vorticity cylindrically polarized vector beams~\cite{salla2015recovering}, and partially coherent radially polarized twisted beams~\cite{zhou2022polarization}. {Although the self-healing of the intensities of vector Bessel beams were demonstrated~\cite{niv2004propagation,he2013propagation}, exploiting the advanced Stokes polarimetry, allowed for precise characterisation and measurement of the vector/ polarization states~\cite{milione2015measuring}}. In addition, the toolkit of classical entanglement provided quantum analogue measures, e.g. the fidelity, concurrence and entropy, to quantitatively describe the self-healing vector beams (Fig.~\ref{f11})~\cite{PhysRevA.98.053818}.

\IN{To see how this self healing holds for nonseparability of vector beams, consider a vector mode}
\begin{equation}
    \ket{\psi} = \frac{1}{\sqrt{2}} \left( a \ket{R}\ket{\ell} + b \ket{L}\ket{-\ell} \right),
    \label{VM}
\end{equation}
\IN{where the right ($R$) and left ($L$) circular polarisation components are entangled to the OAM $\ket{\pm\ell}$ modes while $a$ and $b$ are normalised coefficients. Here the polarisation and OAM components are coupled in a nonseparable way, analogous to quantum entangled systems. The nonseparability between the spatial and polarisation components can be quantified using concurrence (degree of nonseparability) or the vector quality factor  \cite{mclaren2015measuring, ndagano2016beam} which can be measured via state tomography (also see reference \cite{toninelli2019concepts} for experimental  techniques used for reconstructing vector modes). When the field is partially blocked by an opaque obstruction that has no local phase variations and is not bifringent, only diffraction effects will affect the field upon propagation. Furthermore, for simplicity, we assume the obstruction is in the form of a disc that blocks light in the path of the beam. As such the spatial components will diffract radially, i.e., $\ket{\pm\ell} \rightarrow \ket{\pm \ell} \iint c(k_r) \ket{ k_r} dk_r^2$ \cite{sorelli2018diffraction}, where $\ket{k_r}$ are transverse radial modes and $c(k_r)$ are the coefficients. But the scattered radial modes are independent of the azimuthal components and therefore the orthogonality between the spatial modes of the vector mode is preserved. It follows that the nonseparability is also maintained when measured after the shadow region \cite{otte2018recovery} but should still be preserved at any location where the field has nonzero amplitude.} Importantly, the self-healing of classical entanglement enables the quantum-like classical technology of communication and cryptography towards higher security~\cite{otte2020high,https://doi.org/10.1002/lpor.202100533}.

Recently, in addition to the Bessel-type vector beams, other kinds of complex vector beams have been reported, for instance, Airy-type self-accelerating vector beams~\cite{bar2016unveiling} and parabolic-accelerating vector waves~\cite{zhao2021parabolic}. These new types of vector beams combine self-acceleration with the vector patterns of light, which has the potential to produce new forms of self-healing properties in complex vector beams.

\begin{figure}
\centering
\includegraphics[width=0.7\linewidth]{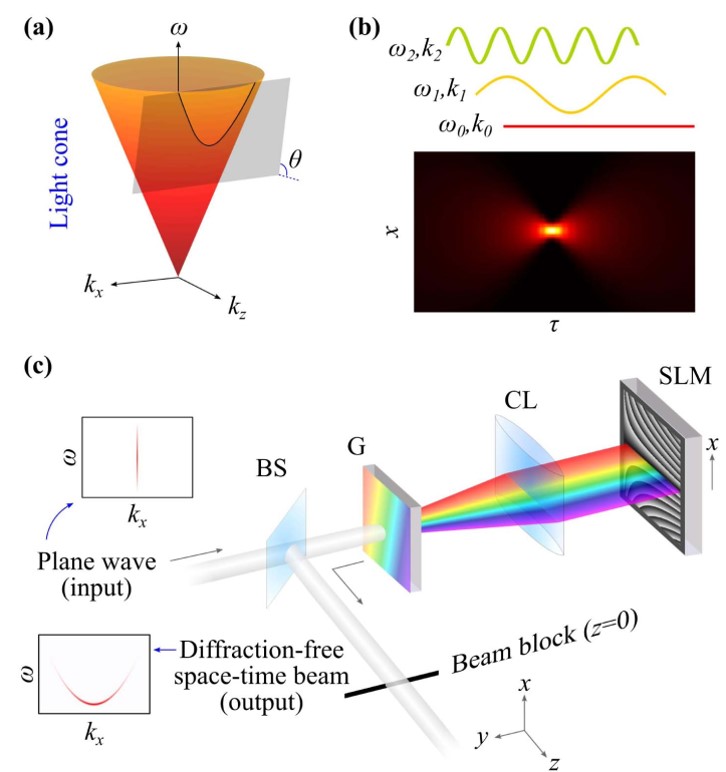}
\caption{\label{st}{Self-healing of space-time light sheets: (a,b) representation of a space-time light sheet (a) on a light cone and (b) in space-time domain; (c) setup to experimentally study the self-healing of space-time light sheets~\cite{kondakci2018self}. \YS{Copyright by Optica Publishing Group} }}
\end{figure}

\subsection{Self-healing in space-time}
Optical modes are usually treated as space-time separable solutions of Maxwell's equations, including all the spatial modes discussed above. In contrast to this conventional viewpoint, the study of spatiotemporal wave packets with space-time nonseparability recently attracted great attention. The space-time nonseparability in spatiotemporal pulses also possess similarity to the space-polarization nonseparability in vector beams, and classical entanglement can be an effective tool to characterize it~\cite{kondakci2019classical,shen2021measures}. A typical example of the space-time pulse is the space-time light-sheet wave packet, in which the 2D pattern in space-time can be nearly nondiffracting upon propagation~\cite{kondakci2017diffraction,yessenov2019weaving}. The self-healing property of space-time light sheets was experimentally verified, whereby the space-time pattern was observed to reconstruct itself upon propagation after a partial obstruction (Fig.~\ref{st})~\cite{kondakci2018self}. In this work, the quantitative characterization of this self-healing process of space-time patterns was also reported. With the rapid development of space-time light modulation, more complex patterns are available to be controlled in space-time, such as Airy wave packets accelerating in space-time~\cite{kondakci2018airy} and space-time caustic light waves~\cite{wong2021propagation}. Therefore, the exploration of self-healing in more kinds of space-time waves may be an intriguing research direction in the near future.

\begin{figure}
\centering
\includegraphics[width=0.7\linewidth]{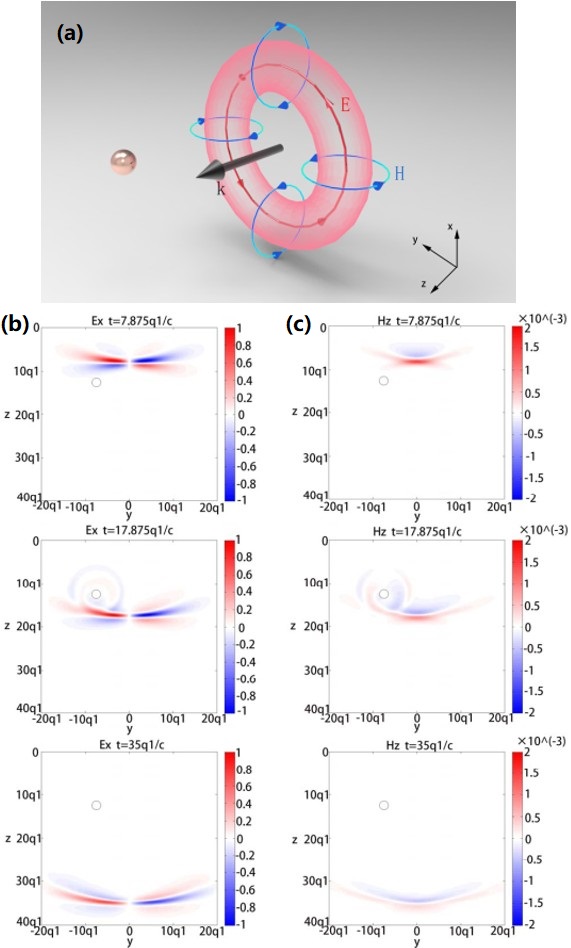}
\caption{\label{f12}{Self-healing of a flying electromagnetic doughnut: (a) Conceptual schematic; (b,c) Simulated results of (b) the electric and (c) magnetic fields of a self-healing flying doughnut~\cite{zhang2019study}. Credit by Ren Wang. }}
\end{figure}

\subsection{Self-healing in space-time vector pulses}
Spatiotemporally structured pulses include not only space-time nonseparable scalar waves, but also more complex space-time-polarization nonseparable vector pulses, which became a recent hot topic. For instance, the space-time vector pulses include picosecond-level space-time vector light sheets~\cite{diouf2021space}, i.e. the space-time light sheets with radially or azimuthally polarized modulation, as well as the vector spatiotemporal vortex that carries transverse vector patterns and transverse OAM~\cite{ChenWanChongZhan+2021+4489+4495}. Recently, the self-healing of the space-time vector light sheets has been experimentally verified and quantitatively characterized~\cite{shen2021measures}.

The space-time vector pulses also include femtosecond and few-cycle level structured pulses. A typical example of these is the toroidal pulses, also named ``flying doughnut'', the theoretically focused single-cycle pulse with toroidal electromagnetic configuration, which were experimentally generated very recently~\cite{zdagkas2021observation}. The toroidal pulses have salient space-time-polarization nonseparability~\cite{shen2021measures}, nondiffracting properties~\cite{shen2022nondiffracting}, and sophisticated spatiotemporal topological 
electromagnetic fields~\cite{shen2021supertoroidal}, providing a platform to study its self-healing properties in the novel space-time vector structures. 
The self-healing of toroidal pulses were theoretically studied recently (Fig.~\ref{f12})~\cite{zhang2019study}, where the space-time toroidal configuration cut by a partial obstruction can repair itself back to a completed toroid upon propagation.

\subsection{Self-healing in waveguides}
In addition to optical modes in free space, the self-healing effects were studied in optical waveguides. With recent developments in optical fabrication, customized structures of photonic waveguides or fibres can be produced, that can realize more complex structured light modes, as well as their amplification ~\cite{wang2021generation, pidishety2017orbital, pidishety2016all,zhu2018multimode, pidishety2019raman}. These structured waveguide modes including the 2D Airy-like or caustic-like modes, the self-healing properties of which were theoretically and experimentally demonstrated~\cite{zhou2019self,deng2013two}. Recently, it was also proved that 1D spatial profiles can also possess the self-healing property in specially designed integrated waveguides~\cite{fang20211d}. Another research direction is to study the self-healing of the temporal profile of structured, pulsed modes in waveguides or fibres. In contrast to the light waves in free-space, the chirped pulses propagating in waveguides can have strong nonlinear interactions with the media, providing a complex mechanism of pulse profile shaping~\cite{agrawal2000nonlinear}. It was found that, in some specially designed waveguide systems, the temporal profile of a truncated structured mode {demonstrates the self-healing effect by evolving into the profile of the original mode. This has been studied in }chirped pulse silicon waveguides~\cite{mandeng2013chirped,mandeng2012nonlinear}, chirped pulse amplifier fibre system~\cite{shen2017gain}, and waveguides under higher-order phase modulations~\cite{banerjee2018self}.

\begin{figure}
\centering
\includegraphics[width=0.9\linewidth]{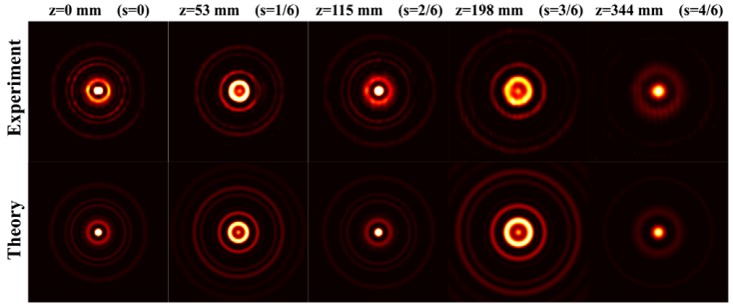}
\caption{\label{f13}{Pattern revivals from fractional Gouy phases in structured light~\cite{da2020pattern}}, \YS{Copyright by American Physical Society}}
\end{figure}

\subsection{Self-healing like effects without an obstruction}
It is understood that structured light may self-heal after certain opaque obstructions, but how about more general nonopaque aberrated obstruction? Recent works verified that the propagation of a traditional self-healing beam through weak aberrations may not guarantee the self-healing, which leads to an interesting conclusion that strong aberrations (e.g. opaque obstructions) may induce self-healing while weak aberrations will not~\cite{mphuthi2018bessel}. However, in some specific conditions, a self-healing analogue effect can also emerge even without an obstruction. For instance, the well-known Talbot effect or self-imaging effect is an effect that the optical pattern of spatial modulation imposed by a diffraction grating would periodically revive itself in the near field pattern~\cite{wen2013talbot}. The Talbot effect can be seen as a pattern self-healing process, but it is conventionally limited in grating near field diffraction. Recently, the Talbot effect was also studied in its counterparts in paraxial structured beams. It was demonstrated that a superposed spatial beam mode fulfilling the fractional Gouy phase condition can reconstruct its initial pattern at the corresponding longitudinal positions (Fig.~\ref{f13})~\cite{da2020pattern}. Studies of Talbot-like effects in optical beams were also recently transferred into the space-time domain, where diverse space-time patterns were designed with the self-reconstructing effect by controlling the space-time coupling in structured pulses~\cite{hall2021space,yessenov2020veiled,hall2021temporal}.

\subsection{Self-healing of other complex structured light}
With the recent rapid development of structured light and OAM, the manipulation of increasingly complex optical modes is moving towards multiple degrees of freedom and higher dimensions, thus the self-healing effect has broader frontier to study. For instance, the self-healing of vortex beams is no longer limited to the simple mode with a phase singularity, but also extended to increasingly complex modes, such as vortex lattices with multiple singularities~\cite{shen2019optical}, helico-conical optical beams~\cite{hermosa2013helico,singh2014conical}, and bored helico-conical beams~\cite{zeng2022self}. Recently, the self-healing properties of many kinds of complex vortex lattices were theoretically and experimentally demonstrated~\cite{zhao2021self}. The self-healing was usually studied in propagating modes, while the study of oscillating modes in a laser resonator is elusive. A recent work proposed a self-healing property in a metasurface laser resonator, which is used to emit topological vortex lattices~\cite{piccardo2021optical}. As for the structured light in waveguides, the matasurface method was also introduced in the novel waveguide design~\cite{meng2021optical}, thus more complex self-healing of structured light in waveguides are to be explored. With the recent emergence of topological photonics and topological photonic crystals, the self-healing was also studied in the modes of photonic crystals, for instance, the self-healing of topological skin modes was demonstrated by engineering the non-Hermitian effect in a photonic crystal system~\cite{PhysRevLett.128.157601}. In addition to the studies of self-healing in propagating modes, a more emerging direction is to study the self-healing of complex structured light oscillating in laser cavities. In a very recent work, a metasurface laser was reported to emit a vortex array, whereby the vortex array oscillation in the cavity demonstrated a self-healing behavior, even if an obstruction was placed on the intracavity metasurface~\cite{piccardo2022vortex}.

\section{Applications of self-healing}
The self-healing property can be a representation of an inertia in optical beams, as it prevents change and {maintains} the shape of optical beams, even after encountering substantial scattering objects within the beam’s path. Such a peculiar property not only allows optical beams to behave robustly, but also enables one to realize {almost-unattainable) applications such as classocal and quantum communication through turbulent atmospheres, microscopy, micromanipulation, ghost imaging, as well as versatile probes for light-matter interactions}, as detailed below.  

\subsection{Optical communication}

Free-space Optical Communication (FSO) has attracted more and more interest over recent years as it provides an easy means of high-bit-rate (BR) communication. The atmospheric turbulence is a critical challenge for FSO links as it causes random fluctuations in amplitude and phase of the optical channel beams. These fluctuations can lead to an increase in the link error probability, and induce crosstalk between multiple channels that limits the performance of communication systems. Self-healing enables optical beams to {maintain} their original shape even after substantial scattering when propagating through a static or non-static random turbulent media.  This benefits FSO in three ways: firstly, since the shape of the optical beam is maintained, longer FSO links can be achieved (without increasing the diameter of the beam significantly allowing adequate power to be received at the detector). Secondly, different structured beams/modes can be used simultaneously as independent communication channels. The third advantage is the potential of significant minimization in random intensity fluctuations at the  detection site lead to an increase in the accuracy of information decoding~\cite{vetter2019realizationPD,wang2018employing}. {Apart from making use of the OAM modes to encode information, the accompanying radial modes associated with LG and BG mode sets have been used~\cite{ahmed2016mode,chen2016demonstration}.} Evidently, in a comparative study of individual channel efficiency between LG and BG beams propagating through turbulence, a large improvement was demonstrated by utilizing BG over LG beams, especially at high levels of turbulence ~\cite{birch2015long, doster2016laguerre}. In the same work, they also showed a drop in channel efficiency with increasing mode numbers and a reduction in {bit-coding} errors when increasing the mode spacing ~\cite{doster2016laguerre}. Although, Bessel beams possessing an inherent self-healing property can reconstruct itself after encountering an obstruction, the strong turbulence will also impair the nondiffracting property causing phase and intensity distortions. The influence of such turbulence can be mitigated, and the self-healing property is recuperated by using adaptive optics-based compensation techniques. 
{With a two Bessel beam multiplexed system, where each beam carried data of 10-Gbaud (40-Gbit/s) 16-ary quadrature amplitude modulation (16-QAM), effective suppression of the inter-channel crosstalk, as well as an improvement in the bit-rate performance and recuperating the nondiffracting property were all experimentally demonstrated(Fig.~\ref{f14}(a))~\cite{li2017adaptive}.} Such FSO links have been studied using different types of beams such as nonuniformly correlated Bessel beams (NUCBs)~\cite{zhang2021propagation}, Airy beams~\cite{nelson2014propagation,zhu2018obstacle,zhu2021free,yong2021propagation,fickler2021quantum}, OAM beams~\cite{singh2022performance,wang2021experimental,willner2021perspectives,wang2022orbital}, and complex multi-vortex beams~\cite{wan2021divergence}. {Since BG beams change from a Bessel structure to an annular ring, this is a disadvantage for long-range communication. This can, however be overcome by implementing long-range Bessel beams which have have longitudinally dependent cone angles and possess self-healing abilities much like usual Bessel beams~\cite{mphuthi2019free,litvin2015self}.} High-dimensional quantum key distribution using hybrid spin-orbit Bessel states~\cite{nape2018self} have been studied in different turbulence conditions, demonstrating that the self-healing property of employed optical channel beams is the key added advantage for the significant improvement in the performance of FSO systems.

\subsection{Microscopy and imaging}

Laser beams possessing the ability to self-reconstruct are crucial for light-sheet fluorescence microscopy (LSFM) imaging, as their initial beam profile can be preserved to a larger extent (even in the presence of massive phase perturbations0, allowing them to propagate deeper into inhomogeneous media (Fig.~\ref{f14}(b))~\cite{fahrbach2012propagation}. As such, they perform well with respect to penetration depth and directional propagation stability in thick biological samples, such as skin or brain tissue or plants, as well as scattering synthetic materials, resulting in a significant increase in image contrast in confocal line microscopy ~\cite{fahrbach2012propagation, fahrbach2010microscopy, tomer2012quantitative, krzic2012multiview, huisken2004optical}. The intrinsic diffractive nature of traditional Gaussian beams limits the non-homogeneous field of view (FOV) as a resulting sheet is strongly degraded as it propagates through the scattering samples ~\cite{tomer2012quantitative, krzic2012multiview, fahrbach2012propagation}. Whereas the self-healing nature of Bessel beams make them ideal to create light-sheets with more extended depth of field (DoF) and homogeneity that results in a more uniform sample illumination, producing wider FOVs, that enables the visualization of large biological samples like mouse brains in LSFM applications~\cite{tomer2012quantitative, krzic2012multiview , fahrbach2012propagation , gao20143d , fahrbach2010microscopy }. \YS{Evidently, a systematic investigation on the self-reconsturction of holographically shaped, scanned Bessel beam by passing it through three different classes of refractive index inhomogeneity, demonstrated in a prototype of a microscope with self-reconstructing beams (MISERB) show that self-reconstructing beams are robust against deflection at objects, that  not only reduces scattering artefacts, but also simultaneously increases image quality and penetration depth in dense media ~\cite{fahrbach2010microscopy}. In Addition,} based on the comparisons in identical experimental conditions, it is deduced that the FOV of a Bessel beam is on average nine times larger than that of a Gaussian beam, without compromising spatial resolution, regardless of the excitation wavelength ~\cite{luna2022multicolor}.  
The main drawback of Bessel beams for LSFM is their side intensity lobes, which illuminate out-of-focus sample regions. It can be reduced by using confocal line-scanning detection, spatial filtering techniques, or beam subtraction methods ~\cite{deng2020subtraction, zhou2020application }. A promising tool to create a thinner light sheets is the optical lattice formed with the coherent Bessel beams in the lattice light-sheet microscopy, where the Bessel beams interfere with each other coherently, and the ring structure is somewhat suppressed ~\cite{ chen2014lattice}. On the other hand, propagation-invariant Airy beams, innately yield a tenfold and fourfold increase in FOV compared to the single-photon Gaussian and Bessel beam light sheets, respectively, while maintaining the high resolution ~\cite{hosny2020planar,  deng2022enhancement, hosny2020planar }. It is also found that the self-healing process is fast in the case of a weak turbid medium, a small size beam width, and low initial coherence, as such this can be used to improve the image quality in a ghost imaging system in a turbid medium (Fig.~\ref{f14}(d))~\cite{zhou2020application, pan2021enhancing }. 

\subsection{Micromanipulation or tweezers}

\begin{figure}
\centering
\includegraphics[width=1.2\linewidth]{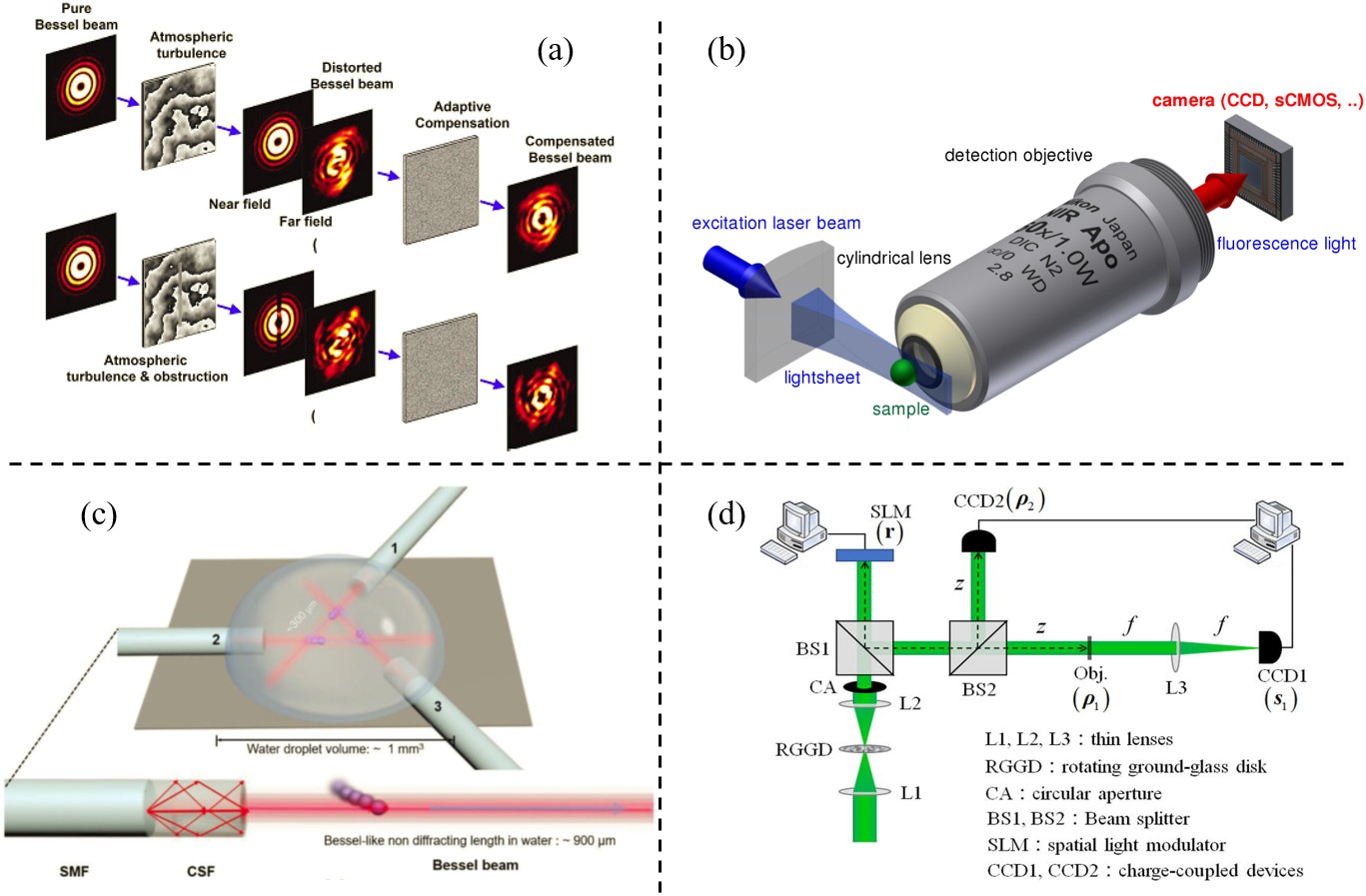}
\caption{\label{f14}{Different applications of self-healing beams: (a): free-space optical communication using Bessel beams~\cite{li2017adaptive}, \YS{Copyright by Spinger Nature}. (b): light-sheet microscopy~\cite{park2021optical}, \YS{Copyright by Wikimedia}. (c): obstacle-circumventing micro-manipulation~\cite{garces2002simultaneous}, \YS{Copyright by Spinger Nature}. (d): ghost imaging with self-healing partially coherent photons \cite{zhou2020application}, \YS{Copyright by AIP Publishing}.}}
\end{figure}

Optical tweezers are a powerful and noninvasive tool for manipulating small objects based on the interaction between light and matter to manipulate micro-objects through momentum transfer. This is indispensable in many fields, including physics, biology, and soft condensed matter~\cite{yang2021optical}. Traditionally, Gaussian light beams are used in optical tweezers, which cannot trap particles in multiple longitudinal locations of more than a few micrometres in separation, as a Gaussian beam possesses a short Rayleigh length from the focal plane and are prone to distortion by the particle. Whereas, the inherent non-diffracting and self-healing nature of Bessel beams allows it to reconstruct itself after a characteristic propagation distance if part of the beam is obstructed or distorted by the trapped particle{~\cite{arlt2001optical}}. With such unique capabilities, Bessel beams are utilized within optical tweezers to trap particles in multiple, spatially separated sample cells with a single beam in the axial direction. These multiple particle tweezers potentially offer enhanced control of ‘lab-on-a-chip’ and optically driven microstructures to carry out simultaneous studies of identically prepared ensembles of colloids and biological matter (Fig.~\ref{f14}(c))~\cite{garces2002simultaneous}. A tweezer that could manipulate particles simultaneously in two different sample cells even with 3mm longitudinal separation, is experimentally demonstrated~\cite{garces2002simultaneous}. Taking the tweezers to a different level, three distinctive motions: the acceleration, the linear motion, and the deceleration have been achieved via controlled circulation of dielectric microparticles over triangular and rectangular polygonal paths formed by crossing multiple Bessel-like beams within a single water droplet ~\cite{park2021optical}. In recent years, rapid progress in tweezers have opened-up novel capabilities in the study of micromanipulation in liquid, air, and vacuum. Which exploits not only the unusual beam properties, such as phase singularities on-axis and propagation-invariant nature, but also different structured light beams with customized phase, amplitude, and polarization in optical trapping ~\cite{yang2021optical,otte2020optical}. Although, the tweezers that exploit the self-healing and nondiffracting characteristics of different optical beams such as LG, Bessel, Mathieu-Gauss, perfect vortex, Airy-Gaussian and Ince-Gaussian beams have taken special position for unique application and pushing the limits of traditional optical tweezers (~\cite{yang2021optical} and references therein). 

\subsection{Self-healing enabled quantum technologies}
We explore two main applications of self-healing in the domain of quantum communication and quantum imaging or microscopy. In these applications, preserving quantum information encoded in photons can be vital for ensuring that high information transmission rates are achieved and hence high dimensional transverse spatial modes are highly attractive \cite{Erhard2018, forbes2019quantum, cozzolino2019high}. Unlike polarisiton, the OAM states span a much larger Hilbert space, therefore promising higher mutual information \cite{Mafu2013A}, robustness against noise \cite{ecker2019overcoming} and increased security against quantum cloning \cite{Bouchard2017}.
While the LG mode family of photons that carry OAM are favoured, they are still sensitive to diffraction \cite{sorelli2018diffraction} and random media \cite{cox2020structured} therefore limiting their performance in practical quantum applications. For this reason, self-healing provides an avenue for overcoming some of these issues in single photon and entangled photon quantum channels.

Some mode families that have been generated at the single photon level include BG \cite{mclaren2013two,baghdasaryan2020characterization } and Airy \cite{lib2020spatially} modes. The BG modes encoded with OAM states have previously been used as a computational basis for information encoding, for applications in QKD. This makes it attractive for increasing key generation rates and security, while self-healing (see tutorial on self-healing quantum key distribution \cite{otte2020high}) is used for overcoming diffraction induced losses due to solid obstructions \cite{nape2018self}. On the other hand, the generation of entangled self accelerating Airy photons \cite{lib2020spatially}, opens up the possibility of using an alternative basis to OAM states. To this end, heralding schemes for Airy photons have been reported \cite{li2021self}, showing that the Airy photon fields can gradually recover their field profile upon propagation. In this demonstration, the photon field was generated from an SPDC source and measured using a single photon sensitive camera.

While robustness to solid obstructions is clearly one main benefit of using self-healing modes, it isn't clear whether there is any robustness against phase dependent aberrations \cite{mphuthi2018bessel}. Although this is true for pure coherent fields, partially coherent fields on the other hand still show resilience in ghost imaging through scattering and turbid media \cite{zhou2020application}. In ghost imaging, one photon interacts with an object and is collected with a detector that has no spatial resolution (bucket detector) while its correlated twin is detected with a spatially sensitive {detector} (we refer the reader to ref. \cite{shapiro2012physics} for more insight on ghost imaging). The detected signals are then correlated and used to reconstruct the object features. The challenge is that the photon signals are sensitive to perturbations (scattering) arising from the environment, and as a result, object information is lost. However, the authors in ref. \cite{zhou2020application} solve this problem by exploiting the self-healing of the degree of spatial coherence, meaning that self-healing can be used to improve the image quality of ghost imaging systems that have low initial coherence or where the photons are corrupted by scattering effects in the propagation medium (Fig.~\ref{f14}(d)).

Although, most significant applications of self-healing of light have been covered here, the exploitation of self-healing of light is increasingly expanding in different fields in the form of different structured beams ~\cite{yang2021optical, willner2021perspectives,otte2020optical }, to carrying out insights into the fundamental studies ~\cite{ liu2022analogous, li2020determining} and for realizing futuristic applications like quantum technologies ~\cite{ mclaren2014self }.

\section{Outlook}
Although the self-healing of light is a venerable topic since its first observation decades ago, it is only recent that we have managed to control and exploit it in diverse forms of structured light. \YS{Although nowadays the classical self-healing effect of light can be fully simulated by wave optics theory, it is still worth studying more elegant and generalized theoretical explanations, such as the ray-wave model, the self-healing in space-time and that of quantum states, as well as to find more practical applications correspondingly.} 
On the other hand, many new kinds of self-healing effects in novel structured light modes are to be explored. This is to a great extent pushed by the recent rapid development of structured light towards multiple degrees of freedom and higher dimensions, allowing optical self-healing effects to be extended from the spatial domain to the space-time domain, and even to the multi-dimensional domain.

\bibliographystyle{spphys}       
\bibliography{sample}   

\end{document}